\documentclass[letterpaper]{article}
\usepackage{graphicx, amssymb}%, showlabels}

\newtheorem{Def}{Def.}[section]
\newtheorem{Thm}[Def]{Theorem}

\newcommand{\spc}{\;\;\;\;\;\;\;\;\;\;}

\newcommand{\bra}{\mbox{$< \!\!$ \nolinebreak}}
\newcommand{\ket}{\mbox{\nolinebreak $>$}}

\newcommand{\R}{\mathbb{R}}
\newcommand{\1}{\mbox{\rm 1 \hspace{-1.05 em} 1}}

\newcommand{\beq}{\begin{equation}}
\newcommand{\eeq}{\end{equation}}

\title{Fermion Systems in Discrete Space-Time}
\author{Felix Finster}
\date{September 2006}

\begin{document}
\maketitle
\begin{abstract}
Fermion systems in discrete space-time are introduced as a
model for physics on the Planck scale. 
We set up a variational principle which describes a non-local interaction
of all fermions.
This variational principle is symmetric
under permutations of the discrete space-time points.
We explain how for minimizers of the variational principle, the
fermions spontaneously break this permutation symmetry
and induce on space-time a discrete causal structure.
\end{abstract}

It is generally believed that the concept of a space-time continuum (like
Minkowski space or a Lorentzian manifold) should be modified for distances as
small as the Planck length. We here propose a concise model where
we assume that space-time is discrete on the Planck scale. Our notion of ``discrete space-time'' differs from other discrete approaches
(like for example lattice gauge theories or spin foam models) in
that we do not assume any structures or relations between the space-time
points (like the nearest-neighbor relation on a lattice or
a causal network).
Instead, we set up a variational principle for an ensemble of quantum mechanical
wave functions. The idea is that for mimimizers of our variational principle,
these wave functions should induce relations between the discrete space-time points, which, in a suitable limit, should go over to the topological and causal structure of a Lorentzian manifold.

The concepts outlined here are worked out in
detail in a recent book~\cite{PFP}. Furthermore, in this book
the connection to the continuum theory is made precise by introducing the
notion of the {\em{continuum limit}}, and mathematical methods are
developed for analyzing our variational principle in this limit.
More specifically, in the continuum limit the fermionic wave functions
group to a configuration of Dirac seas; for details see~\cite{reg}.
Analyzing our variational principle in the continuum limit
gives concrete results for the effective continuum theory;
see~\cite{PFP} and the review article~\cite{rev}.

In this short article we cannot enter the
constructions leading to the continuum limit. Instead, we
introduce the mathematical framework in the discrete setting
(Sections~\ref{sec1} and~\ref{sec1a}) and discuss it afterwards, working out the underlying physical principles (Section~\ref{sec2}).
We finally describe the spontaneous symmetry breaking and the
appearance of a ``discrete causal structure'' (Section~\ref{sec3}).

\section{Fermion Systems in Discrete Space-Time} \label{sec1}
We let~$(H, \bra .|. \ket)$ be a complex inner product space of signature~$(N,N)$.
Thus~$\bra .|. \ket$ is linear
in its second and anti-linear in its first argument, and it is symmetric,
\[ \overline{\bra \Psi \:|\: \Phi \ket} \;=\; \bra \Phi \:|\: \Psi \ket \quad
\spc {\mbox{for all~$\Psi,\Phi \in H$}} \,, \]
and non-degenerate,
\[ \bra \Psi \:|\: \Phi \ket \;=\; 0 \;\;\; {\mbox{for all $\Phi \in H$}}
 \quad \Longrightarrow \quad
\Psi \;=\; 0 \:. \]
In contrast to a scalar product, $\bra .|. \ket$ is {\em{not}} positive.
Instead, we can choose an orthogonal
basis~$(e_i)_{i=1,\ldots,2N}$ of~$H$ such that the inner product
$\bra e_i \,|\, e_i \ket$ equals~$+1$ if $i=1,\ldots,N$ and equals~$-1$ if
$i=N+1,\ldots,2N$.

A {\em{projector}}~$A$ in~$H$ is defined just as in Hilbert spaces as a linear
operator which is idempotent and self-adjoint,
\[ A^2 = A \spc {\mbox{and}} \spc \bra A\Psi \:|\: \Phi \ket = \bra \Psi \:|\: A\Phi \ket \quad
{\mbox{for all $\Psi, \Phi \in H$}}\:. \]
Let~$M$ be a finite set. To every point~$x \in M$ we associate a projector
$E_x$. We assume that these projectors are orthogonal and
complete in the sense that
\beq \label{oc}
E_x\:E_y \;=\; \delta_{xy}\:E_x \spc {\mbox{and}} \spc
\sum_{x \in M} E_x \;=\; \1\:.
\eeq
Furthermore, we assume that the images~$E_x(H) \subset H$ of these
projectors are non-de\-ge\-ne\-rate subspaces of~$H$, which
all have the same signature~$(n,n)$.
We refer to~$(n,n)$ as the {\em{spin dimension}}.
The points~$x \in M$ are
called {\em{discrete space-time points}}, and the corresponding
projectors~$E_x$ are the {\em{space-time projectors}}. The
structure~$(H, \bra .|. \ket, (E_x)_{x \in M})$ is
called {\em{discrete space-time}}.
A space-time projector~$E_x$ can be used to project vectors of~$H$
to the subspace~$E_x(H) \subset H$. Using a more graphic notion, we
also refer to this projection as the {\em{localization}} at the
space-time point~$x$.

In order to describe the particles of our system, we introduce one
more projector~$P$ in~$H$, the so-called {\em{fermionic projector}}, which
has the additional property that its image~$P(H)$ is a
{\em{negative definite}} subspace of~$H$.
The vectors in the image of~$P$ have the interpretation as the
occupied fermionic states of our system, and thus the rank of~$P$
gives the {\em{number of particles}} $f := \dim P(H)$.

We call the obtained system~$(H, \bra .|. \ket, (E_x)_{x \in M}, P)$
a {\em{fermion system in discrete space-time}}.
Note that our definitions involved only
three integer parameters: the spin dimension~$n$, the number
of space-time points~$m$, and the number of particles~$f$.

\section{A Variational Principle} \label{sec1a}
In order to introduce an interaction of the fermions, we
shall now set up a variational principle.
To this end, we need to form composite
expressions in our projectors~$(E_x)_{x \in M}$ and~$P$.
It is convenient to use the short notations
\beq \label{notation}
P(x,y) \;=\; E_x\,P\,E_y \spc {\mbox{and}} \spc
\Psi(x) \;=\; E_x\,\Psi \:.
\eeq
The operator~$P(x,y)$ maps~$E_y(H) \subset H$ to~$E_x(H)$, and it is often
useful to regard it as a mapping only between these subspaces,
\[ P(x,y)\;:\; E_y(H) \: \rightarrow\: E_x(H)\:. \]
Using~(\ref{oc}), we can write the vector~$P \Psi$ as follows,
\[ (P\Psi)(x) \;=\; E_x\: P \Psi \;=\; \sum_{y \in M} E_x\,P\,E_y\:\Psi
\;=\; \sum_{y \in M} (E_x\,P\,E_y)\:(E_y\,\Psi) \:, \]
and thus
\beq \label{diskernel}
(P\Psi)(x) \;=\; \sum_{y \in M} P(x,y)\: \Psi(y)\:.
\eeq
This relation resembles the representation of an operator with an integral kernel, and therefore we call~$P(x,y)$ the {\em{discrete kernel}} of the fermionic projector. Next we define the {\em{closed chain}}~$A_{xy}$ by
\beq \label{cc}
A_{xy} \;=\; P(x,y)\: P(y,x) \;=\; E_x \:P\: E_y \:P\: E_x \:;
\eeq
it maps~$E_x(H)$ to itself.
Let~$\lambda_1,\ldots,\lambda_{2n}$ be the zeros of the characteristic polynomial
of~$A_{xy}$, counted with multiplicities. We define the {\em{spectral weight}}~$|A_{xy}|$
by
\[ |A_{xy}| \;=\; \sum_{j=1}^{2n} |\lambda_j|\:. \]
Similarly, one can take the spectral weight of powers of~$A_{xy}$, and by summing
over the space-time points we get positive numbers depending only on the
form of the fermionic projector relative to the space-time projectors.
Our variational principle is to
\beq \label{vary}
{\mbox{minimize}} \quad \sum_{x,y \in M} |A_{xy}^2|
\eeq
by considering variations of the fermionic projector which satisfy the constraint
\beq \label{constraint}
\sum_{x,y \in M} |A_{xy}|^2 = {\mbox{const}} \:.
\eeq
In the variation we also keep the number of particles~$f$ as well as
discrete space-time fixed.
Using the method of Lagrange multipliers,
for every minimizer~$P$ there is a real parameter~$\mu$ such that~$P$
is a stationary point of the {\em{action}}
\beq \label{Sdef}
{\mathcal{S}}_\mu[P] \;=\; \sum_{x,y \in M} {\mathcal{L}}_\mu[A_{xy}]
\eeq
with the {\em{Lagrangian}}
\beq \label{Ldef}
{\mathcal{L}}_\mu[A] \;=\; |A^2| - \mu\: |A|^2 \:.
\eeq

This variational principle was first introduced in~\cite{PFP}. In~\cite{F1}
it is analyzed mathematically, and it is shown in particular that
minimizers exist:
\begin{Thm} \label{thmn1}
The variational principle~(\ref{vary}, \ref{constraint}) attains its minimum.
\end{Thm}
In~\cite[Section~3]{F1} the variational principle is also illustrated
in simple examples.

\section{Discussion of the Underlying Physical Principles} \label{sec2}
We come to the physical discussion. Obviously, our mathematical framework does not
refer to an underlying space-time continuum, and our variational principle is set up
intrinsically in discrete space-time. In other words, our approach is {\em{background free}}.
Furthermore, the following physical principles are respected, in a sense we briefly explain.
\begin{itemize}
\item The {\bf{Pauli Exclusion Principle}}:
We interpret the vectors in the image of~$P$ as the quantum mechanical states of the particles
of our system. Thus, choosing a basis~$\Psi_1,\ldots, \Psi_f
\in P(H)$, the~$\Psi_i$ can be thought of as the wave functions of the occupied states
of the system.
Every vector $\Psi \in H$
either lies in the image of $P$ or it does not.
Via these two conditions, the fermionic projector encodes for every state
$\Psi$ the occupation numbers $1$ and $0$, respectively, but it is
impossible to describe higher occupation numbers.
More technically, we can form the anti-symmetric many-particle wave function
\[ \Psi \;=\; \Psi_1 \wedge \cdots \wedge \Psi_f \:. \]
Due to the anti-symmetrization, this definition of~$\Psi$ is (up to a
normalization constant) independent of the choice of the basis
$\Psi_1,\ldots, \Psi_f$.
In this way, we can associate to every fermionic projector a fermionic
many-particle wave function which obeys the Pauli Exclusion Principle.
For a detailed discussion we refer to~\cite[\S3.2]{PFP}.

\item A {\bf{local gauge principle}}:
Exactly as in Hilbert spaces, a linear operator~$U$ in~$H$ is called
{\em{unitary}} if
\[ \bra U \Psi \:|\: U \Phi \ket \;=\; \bra \Psi \:|\: \Phi \ket
\spc {\mbox{for all $\Psi, \Phi \in H$}}. \]
It is a simple observation that a joint unitary transformation
of all projectors,
\beq \label{unit}
E_x \;\to\; U E_x U^{-1} \:, \qquad
P \;\to\; U P U^{-1} \spc {\mbox{with~$U$
unitary}}
\eeq
keeps our action~(\ref{vary}) as well as the constraint~(\ref{constraint})
unchanged, because
\begin{eqnarray*}
P(x,y) &\to& U\:P(x,y)\:U^{-1} \:,\spc
A_{xy} \;\to\; U A_{xy} U^{-1} \\
\det (A_{xy} - \lambda \1) &\to&
\det \!\left( U (A_{xy} - \lambda\1)\: U^{-1} \right) \;=\;
\det (A_{xy} - \lambda\1)\:,
\end{eqnarray*}
and so the~$\lambda_j$ stay the same. Such unitary transformations can
be used to vary the fermionic projector. However, since we want to keep discrete
space-time fixed, we are only allowed to consider unitary
transformations which do not change the space-time projectors,
\beq \label{gauge1}
E_x \;=\; U E_x U^{-1} \spc {\mbox{for all $x \in M$}}\:.
\eeq
Then~(\ref{unit}) reduces to the transformation of the fermionic projector
\beq \label{gauge2}
P \;\to\; U P U^{-1}\:.
\eeq
The conditions~(\ref{gauge1}) mean that~$U$ maps every subspace~$E_x(H)$ into
itself. Hence~$U$ splits into a direct sum of unitary transformations
\beq \label{local}
U(x) \;:=\; U E_x \;:\; E_x(H) \: \rightarrow\: E_x(H) \:,
\eeq
which act ``locally'' on the subspaces associated to the individual space-time
points.

Unitary transformations of the form~(\ref{gauge1}, \ref{gauge2}) can be identified
with local gauge transformations. Namely, using the notation~(\ref{notation}),
such a unitary transformation~$U$ acts on a vector~$\Psi \in H$ as
\[ \Psi(x) \;\longrightarrow\; U(x)\: \Psi(x)\:. \]
This formula coincides with the well-known transformation law of wave functions
under local gauge transformations (for more details see~\cite[\S1.5 and \S3.1]{PFP}).
We refer to the group of all unitary transformations of the form~(\ref{gauge1}, \ref{gauge2})
as the {\em{gauge group}}. The above argument shows that our variational
principle is {\em{gauge invariant}}.
Localizing the gauge transformations according to~(\ref{local}), we obtain
at any space-time point~$x$ the so-called {\em{local gauge group}}.
The local gauge group is the group of
isometries of~$E_x(H)$ and can thus be identified with the group~$U(n,n)$.
Note that in our setting the local gauge group cannot be chosen arbitrarily,
but it is completely determined by the spin dimension.
\item The {\bf{equivalence principle}}:
At first sight it might seem impossible to speak of the equivalence
principle without having the usual space-time continuum. What we
mean is the following more general notion. The equivalence principle
can be expressed by the invariance of the physical equations under
general coordinate transformations. In our setting, it makes no
sense to speak of coordinate transformations nor of the
diffeomorphism group because we have no topology on the space-time
points. But instead, we can take the largest group which can act on
the space-time points: the group of all permutations of~$M$. Our
variational principle is obviously {\em{invariant under
permutations}} of~$M$ because permuting the space-time points merely
corresponds to reordering the summands in~(\ref{vary},
\ref{constraint}). Since on a Lorentzian manifold, every
diffeomorphism is bijective and can thus be regarded as a
permutation of the space-time points, the invariance of our
variational principle under permutations can be considered
as a generalization of the equivalence principle.
\end{itemize}
Clearly, the permutation symmetry is not compatible with the topological and causal structure of a Lorentzian manifold.
Also, at fist sight it might seem problematic that our definitions involve
{\em{no locality}} and {\em{no causality}}.
We do not consider these principles as being fundamental.
Instead, our concept is that the minimizer~$P$ of our variational principle
should spontaneously break the above permutation symmetry
and should induce a causal structure on the space-time points.
This will be outlined in the next section.

\section{Spontaneous Generation of a Discrete Causal Structure} \label{sec3}
The symmetries of a fermion system in discrete space-time
can be described abstractly working with unitary representations of finite groups in indefinite inner product spaces.
This abstract framework is developed in~\cite{F2}. We here state
one result and discuss it afterwards. As explained above,
discrete space-time~$(H, \bra .|. \ket, (E_x)_{x \in M})$ as well as
our variational principle are symmetric under permutations of the
space-time points.
However, the fermionic projector might destroy this
symmetry. The next definition makes precise what we mean by a
permutation symmetry of the whole system.

\begin{Def} \label{defouter}
A fermion system in discrete
space-time~$(H, \bra .|. \ket, (E_x)_{x \in M}, P)$
is called {\bf{permutation symmetric}} if for every
permutation~$\sigma$ of the space-time points~$M$ there is a unitary
transformation~$U$ on~$H$ such that
\[ UPU^{-1} \;=\; P \qquad {\mbox{and}} \qquad
U E_x U^{-1} \;=\; E_{\sigma(x)} \quad
{\mbox{for all~$x \in M$}}\:. \]
\end{Def}
The following theorem is proven in~\cite{F2}.
\begin{Thm}
Suppose that~$(H, \bra .|. \ket, (E_x)_{x \in M}, P)$
is a fermion system in discrete space-time of spin dimension~$(n,n)$.
If the number of space-time points~$m$ is sufficiently large
and the number of particles~$f$ lies in the range
\[ n \;<\; f \;<\; m-1\:, \]
then the system is not permutation symmetric.
\end{Thm}
Applied in the physically interesting case~$n \ll f \ll m$,
where the number of particles is much larger than the spin dimension
and much smaller than the number of space-time points, this theorem shows that
the fermion system is necessarily less symmetric than discrete space-time
and our variational principle. In other words, the fermionic projector
spontaneously breaks the permutation symmetry.
This result can be understood intuitively as follows.
One method for building up a fermion system with permutation symmetry
would be to localize one or several particles at each space-time point.
But for this we would need at least~$m$ particles, in contradiction
to our hypothesis~$f < m$. Another method would be to work with
fermionic states which have permutation symmetry. Such fermions would
be ``completely delocalized'' in the sense that after knowing the wave
function at one space-time point, we can recover it at any other
space-time point by applying the permutation group.
The orthogonality of such permutation symmetric states means that
these states must even be orthogonal at each space-time point.
Since in addition the states are negative definite, we conclude that
there are at most~$n$ such permutation symmetric states, not enough
to take into account all~$f>n$ particles of our system.
Using results of the representation theory of finite groups,
it is shown that there is indeed no other method for building
up fermion systems with permutation symmetry.

The spontaneous breaking of the permutation symmetry implies that
the fermionic projector induces a non-trivial relations between the
space-time points. The mathematical structure of our
variational principle gives us some insight into the
nature of these relations.
The basic mechanism becomes clear already in
the simplest possible case of spin dimension~$(1,1)$ and the
value~$\mu=1/2$ of the Lagrange multiplier in~(\ref{Ldef}).
In this case, the closed chain~$A_{xy}$ is a $2 \times 2$-matrix;
we denote the zeros of its characteristic polynomial by~$\lambda_\pm$.
Then the Lagrangian~(\ref{Ldef}) becomes
\[ {\mathcal{L}}[A] \;=\; |A^2| - \frac{1}{2} \:|A|^2 \;=\;
\Big( |\lambda_+|^2 + |\lambda_-|^2 \Big) - \frac{1}{2}\:
\Big( |\lambda_+| + |\lambda_-| \Big)^2 \:, \]
and this can be written as
\beq \label{Lsimp}
{\mathcal{L}}[A] \;=\;
\frac{1}{2} \Big( |\lambda_+| - |\lambda_-| \Big)^2 \:.
\eeq
Now we have a good intuitive understanding of our variational principle:
it tries to achieve that the absolute values of~$\lambda_+$ and~$\lambda_-$
are equal.

The closed chain~$A_{xy}$ is a self-adjoint operator on the
indefinite inner product space $(H, \bra .|. \ket)$.
If~$H$ were a Hilbert space, we could conclude that~$A_{xy}$ is
diagonalizable with real eigenvalues. However, this result is
not true in indefinite inner product spaces. In general,
$A_{xy}$ will not be diagonalizable, and even if it is,
its eigenvalues will in general not be real. But at least we know that
the characteristic polynomial of~$A_{xy}$ is real, and this implies
that its zeros~$\lambda_\pm$ are either both real,
or else they must form a complex conjugate pair
(i.e.\ $\overline{\lambda_+}=\lambda_- \not \in \R$).
These two cases allow us to introduce a notion of causality.
\begin{Def}
Two discrete space-time points~$x,y \in M$ are called {\bf{timelike}}
separated if the zeros~$\lambda_\pm$ of the characteristic polynomial
of~$A_{xy}$ are both real. Conversely, they are said to be
{\bf{spacelike}} separated if the~$\lambda_\pm$ form a complex
conjugate pair.
\end{Def}
This definition really reflects the structure of the Lagrangian~(\ref{Lsimp}).
Namely, if~$x$ and~$y$ are space-like separated, the zeros~$\lambda_\pm$
of the characteristic polynomial of~$A_{xy}$ form a complex conjugate
pair. Hence~$|\lambda_+| = |\overline{\lambda_-}| = |\lambda_-|$,
and thus the Lagrangian~(\ref{Lsimp}) vanishes identically.
Computing first variations of the Lagrangian, one sees that
these also vanish, and thus~$P(x,y)$ does not enter the Euler-Lagrange equations. This can be seen in analogy to the usual notion of causality
in Minkowski space that points with space-like separation 
cannot infuence each other.

According to~(\ref{Lsimp}), our variational principle tries to achieve
that as many pairs of points~$(x,y)$ as possible are space-like separated.
On the other hand, the fact that~$P$ projects onto a definite subspace
of~$H$ implies that not all pairs of points can be space-like separated
(this is made precise by the lower bounds of the action
in~\cite[Section~4]{F1}). Hence we can say that the mathematical
structure of our fermion system ensures that certain
space-time points will be timelike separated. On the other hand,
the variational principle favors spacelike separation.
This gives rise to a spontaneous generation of an interesting
causal structure.

%\section*{References}

\noindent
NWF I -- Mathematik,
Universit{\"a}t Regensburg, 93040 Regensburg, Germany, \\
{\tt{Felix.Finster@mathematik.uni-regensburg.de}}


\begin{thebibliography}{99}
\bibitem{PFP} F.\ Finster, ``The Principle of the Fermionic Projector,''
{\em{AMS/IP Studies in Advanced Mathematics}} {\bf{35}} (2006)
\bibitem{F1} F.\ Finster, ``A variational principle in discrete space-time -- existence of minimizers,'' math-ph/0503069, {\em{Calc.\ Var.\ and Partial Diff.\ Eq.}}\
{\bf{29}} (2007) 431-453
\bibitem{F2} F.\ Finster, ``Fermion systems in discrete space-time -- outer symmetries and spontaneous symmetry breaking,'' math-ph/0601039,
{\em{Adv.\ Theor.\ Math.\ Phys.}}\ {\bf{11}} (2007) 91-146
\bibitem{rev} F.\ Finster, ``The Principle of the Fermionic Projector: An Approach for Quantum Gravity?'', gr-qc/0601128,
in ``Quantum Gravity'', B.\ Fauser, J.\ Tolksdorf and E.\ Zeidler, Eds,
Birkh{\"a}user (2006)
\bibitem{reg} F.\ Finster, ``On the regularized fermionic projector of the vacuum,'' math-ph/0612003 (2006)
\end{thebibliography}
\end{document}